\documentclass[runningheads]{llncs}
\usepackage{graphicx}
\graphicspath{ {images/} }
\usepackage{todonotes}
\usepackage{float}
\usepackage{subcaption}
\usepackage[normalem]{ulem} %sout
\usepackage{mathtools}
\newcommand\secL[1]{\label{sec:#1}}
\newcommand\secR[1]{Section~\ref{sec:#1}}

\begin{document}
\title{AMoDSim: An Efficient and Modular Simulation Framework for Autonomous Mobility on Demand}
\titlerunning{AMoDSim: An Efficient and Modular Simulation Framework}
% If the paper title is too long for the running head, you can set
% an abbreviated paper title here
%
\author{Andrea Di Maria\inst{1} \and
Andrea Araldo\inst{2}\orcidID{0000-0002-5448-6646} \and \\
Giovanni Morana\inst{1} \and 
Antonella Di Stefano \inst{3}
}
\authorrunning{Di Maria et al.}
% First names are abbreviated in the running head.
% If there are more than two authors, 'et al.' is used.
%
\institute{Aucta Cognitio R\&D Labs, Catania 95123,  Italy \\
\email{\{adimaria,gmorana\}@auctacognitio.net}\\
\and
R\'eseaux et Services de T\'el\'ecommunications, Te\'l\'ecom SudParis, Evry 91011, France\\
\email{andrea.araldo@telecom-sudparis.eu}\\
\and
Universit\'a di Catania, Catania 95125,  Italy \\
\email{ad@dieei.unict.it}\\
}
\maketitle              % typeset the header of the contribution
\begin{abstract}

Urban transportation of next decade is expected to be disrupted by Autonomous Mobility on Demand (AMoD): AMoD providers will collect ride requests from users and will dispatch a fleet of autonomous vehicles to satisfy requests in the most efficient way. Differently from current ride sharing systems, in which driver behavior has a clear impact on the system, AMoD systems will be exclusively determined by the dispatching logic. As a consequence, a recent interest in the Operations Research and Computer Science communities has focused on this control logic. The new propositions and methodologies are generally evaluated via simulation. Unfortunately, there is no simulation platform that has emerged as reference, with the consequence that each author uses her own custom-made simulator, applicable only in her specific study, with no aim of generalization and without public release. This slows down the progress in the area as researchers cannot build on each other's work and cannot share, reproduce and verify the results.
The goal of this paper is to present AMoDSim, an open-source simulation platform aimed to fill this gap and accelerate research in future ride sharing systems.

%To the best of our knowledge, AMoDSim is the only platform that allows the researcher to easily and rapidly evaluate different algorithms for ride sharing. Therefore, 

\end{abstract}
\section{Introduction}
\secL{introduction}
Transportation is traversing a period of big transformations driven by Information and Communication Technology (ICT). For instance, the ubiquitous connectivity guaranteed by 3G and 4G has triggered the emergence of \emph{ride sharing} services, e.g., Uber and Lyft, in which users reserve a ride through a smartphone app and service providers match them to a fleet of vehicles. Goldman Sachs quantifies the importance of these services by predicting a market of 285 billion dollars in 2030 \cite{Burgstaller2017}. In more and more cities, ride sharing services are also determining a transformation of every-day life~\cite{Clewlow2017}. This revolution will become even deeper when these services will be provided by Autonomous Vehicles (AVs). Autonomous Mobility on Demand (AMoD) services~\cite{Araldo2018} will be very cheap for the users, since providers will not have to sustain the cost of labor of the drivers.

One reason for the efficiency of these systems is that vehicles can be shared among many users. To do so, efficient and scalable algorithms are needed. While the Vehicle Routing Problem \cite{VehileRouting2014} has been studied from the 1950s, the success of ride sharing systems has lead to a renovated interest in this decade, where the problem has been specialized to the case of matching ride requests from passengers to available vehicles, while respecting some constraints on users' waiting and riding time. A particular focus has regarded the computation of condensed vehicle trips to properly aggregate many rides in order to minimize provider's costs while keeping the user quality of level acceptable.
The request-vehicle matching problem has been shown to be NP hard \cite{Alonso-Mora2017}. Therefore, a vast literature has developed to propose ``good'' heuristics with a reasonable computation time to be used in practice and has resorted to simulation to evaluate them. Unfortunately, up to now no reference simulation tool has emerged for this, which is shown by the fact that most of the authors have been forced to build from scratch their own case-specific simulator. The negative consequences are:
\begin{itemize}
\item Waste of time and effort, to create every time a simulator.
\item Impossibility to build on the effort of past research.
\item Difficulty for the community to reproduce and verify results.
\end{itemize}

On the other side, there are few exceptions of complex transportation simulation tools extended with models of ride sharing systems. However, they are not suitable for the researchers interested in the development of algorithms for ride-sharing, whom we target in this work. The reasons are:
\begin{itemize}
\item They require to specify scenarios with high level of realism, like economic indicators of the population and of the area, which are not usually available.
\item Even if available, it takes a long time and effort to figure out how to set them up into the simulators, which would instead be preferable to spend in the inner workings of the algorithms.
\item They lack flexibility: when developing an algorithm, it is necessary to test it in a vast range of scenarios, instead of just super-realistic one, to generalize the findings.
\item The level of detail transportation represents an overhead: part of the computation time is spent in representing the detailed movement of vehicles at millisecond scale, which has no big impact on the ride sharing logic.
\end{itemize}
For these reasons, transportation simulation tools are to be used a-posteriori when, for instance, a transportation authority or company wants to check what is the impact of a ride sharing strategy, already developed and thoroughly studied, on the particular scenario of interest.

In this paper we present AMoDSim, a simulation framework open to  researchers in future-generation ride-sharing systems whose design goals are:
\begin{itemize}
\item Launching massive simulation campaigns to simultaneously test the performance of the algorithms under study, under different settings, is easy and scalable.
\item By means of modularity, it is easy to implement new algorithms, with minimum modification of the other components.
\item Results on the performance for both the provider and the user perspective are produced automatically and are simple to analyze.
\end{itemize}

The code is available\footnote{\url{https://github.com/admaria/AMoDSim} } under the \emph{CC BY-NC-SA 4.0} license. The rest of the paper is organized as follows: In \secR{related} we review the work in simulation of ride sharing systems. In \secR{mod} we present the model of AMoD used in  AMoDSim. In \secR{simulator} we describe its architecture and in \secR{case-study} we showcase it in a case study in which we compare several provider and user-related metrics of two different matching algorithms.

\section{Related Work}
\secL{related}
In this section we describe the state of the art of the research on autonomous mobility on demand and future generation ride sharing systems, focusing on the simulation tools used. We divide this research in works that use case-specific simulators and complex transportation simulators. The limitations of both has been discussed in the previous section.
\subsection{Work Based on Case-Specific Simulators}
We emphasize that no code has been made public with any of the studies listed in this subsection, nor the simulators have been described enough to be reproducible. This reinforces the utility of our effort.
Santi, Frazzoli et Al. published a series of papers \cite{Santi2014,Alonso-Mora2017,Santi2018} where they proposed mathematical formulations of ride sharing problems and heuristics to solve them. Case studies are shown in New York. Similarly, Ma et Al. \cite{Ma2013a} study ride-sharing algorithms using GPS taxi trajectories collected in Bejing.
Agatz et Al. \cite{Agatz2011} built a simulator for a case study in Atlanta. Within their simulator, an agent can subscribe to a provider either as a rider or a driver. The study better represents systems like BlaBla Car \cite{BlaBlaCar}, in which a traveler can publish her future trip in a web portal and other users can hop-in. These systems are now called ``carpooling'' and are different from ride sharing systems like Uber and Lyft and the future AMoD, in which (i) drivers are continuously operating for hours just to serve other individuals' trips and (ii) requests for rides arrive continuously in real time and are not announced in advance. 
Other case-specific simulators were developed for case studies in Seul and Boston in \cite{Jung2013} and \cite{Lam2016}, respectively. 
\subsection{Work Based on Complex Transportation Simulators}
Some case studies have been performed extending commercial transportation simulators, like Aimsun \cite{Correia2015,Linares2016}. However, commercial tools are usually not available to researchers and their code is closed, impeding the verification and the reproduction of results.
To the best of our knowledge, three simulation tools developed by academic institutions have been extended and employed in studies related to AMoD, namely SimMobility \cite{Araldo2018} and MATSim \cite{Boesch2016,Bischoff2016} and SUMO\cite{Alazzawi2018}. The main issue with the first two is the level of complexity that the researcher is required to handle and the performance. They are agent-based, i.e., they simulate the behavior of each single traveler through transportation-specific economic models. In order to do so, the researcher must construct first a synthetic population and describe the economic indicators of the urban network. As discussed in \secR{introduction}, this is overkill for research focused on algorithms, which is what we target here. The unsuitability of these tools is testified by the fact that: (i) they are generally used, at least as far as published research visible to us is concerned, only by the very same group that developed them and (ii) researchers have preferred to craft their own case-specific simulators instead of using them. SUMO is a microscopic simulator that has been employed in a recent case study on AMoD in the city of Milan\cite{Alazzawi2018}. However, that study does not fill the gap we aim to fill. First, SUMO is a purely microscopic simulator, i.e., it computes the detailed movement of each vehicle,\footnote{A particular version of SUMO, called SUMO MESO\cite{SumoMeso}, is intended to reduce the details in vehicle movement simulation. However, we are not aware of any published study on AMoD systems based on SUMO MESO.} 
which is an overhead that we want instead to avoid, since it has limited interest when studying the dispatching logic in an AMoD system. Second, SUMO does support natively Mobility on Demand services and the authors of \cite{Alazzawi2018} had to write from scratch this functionality, which, however, they do not make publicly available. Third, SUMO needs detailed input, that the authors needed to obtain by cross-correlating several data-sources (Google APIs, mobile phone traces, etc.), while the choice we made in AMoD is to streamline the input definition, sacrificing some realism. Finally, is it not possible in \cite{Alazzawi2018} to specify user-specified quality of service requirements.

\subsection{Other work}
NOT IN THIS DRAFT

\section{Model of Autonomous Mobility on Demand}
\label{sec:mod}
We now present the model of AMoD service implemented into the simulator. The model includes a fleet of \emph{vehicles}, a \emph{coordinator} managing it and \emph{users}.
Users send trip requests to the coordinator, which runs matching algorithms or simply orchestrates the distributed computation running in the vehicles, in order to decide how to match them to the available vehicles.
A trip request consists of two \textit{stop}-points, one for the pick-up and one for the drop-off. Each stop point is a tuple $sp = \{q, t, \Delta t\}$, where $q$ is the pick-up or drop-off point, $t$ is the \emph{preferred time} at which the user wishes to be picked up or dropped off, $\Delta t$ is the \emph{maximum} \emph{extra-time} the user tolerates to be picked up or dropped off, with respect to the preferred time.

At any given time, each vehicle $v$ has a set of planned \textit{stop}-points organized in a certain sequence $S_v = [sp_1,sp_2,\dots ]$, that we call \emph{schedule}. Each schedule is associated with a \emph{cost} $c(S_v)$, which can be defined in different ways to take into account provider or user-related metrics. For example, this cost could be the kilometers traveled to accomplish that schedule, or some indication of the travel or waiting time of the users served by that schedule. The goal of the provider is to create and continuously update the schedule $S_v$ of each vehicle of its fleet, in order to optimize the costs $c(S_v)$, subject to respecting the time constraints of all the users. Observe that this model is general enough to represent different types of optimization: (i) both provider cost or user level of service can be optimized, as this boils down to the way the cost $c(S_v)$ is defined; (ii) one can simply study the overall cost optimization, or min-max optimization, etc.; (iii) the optimization can be both centralized, in case a single coordinator decides all the schedules $S_v$, or distributed, in case, for instance, each vehicle $v$ optimizes its own schedule. While the model is general, we have currently only implemented the strategies described in \secR{strategies}.

\subsection{Time constraints}
\label{sec:time}
We define a schedule $S_v=[sp_0,\dots,sp_n]$ of a vehicle $v$ \emph{feasible}, if the time constraints of all its stop-points is satisfied. Let us suppose $sp_i=(q_i, t_i, \Delta t_i)$ and that $b_i$ is the time needed to complete $sp_i$, i.e., the time for the passenger to board (alight), in case of pick-up (drop-off), that the current time is $t_\textnormal{now}$ and the current vehicle location is $q_v$. Let us denote with $\tau(q,q')$ the estimated time to go from a location $q$ to $q'$. Then the estimated time at which the stop-point $sp_i$ will be served is:
\[
	\hat t_i = 
    	t_\textnormal{now}+\tau(q_v,q_0)+
        \sum_{j=1}^{i} 
        	[ b_{j-1}+\tau(q_{j-1},q_j) ] + b_i
\]

The \emph{estimated delay} of each stop-point $d_i = \hat t_i - t_i \le \Delta t_i$, for $i=0,\dots,n$. The provider must only compute feasible schedules $S_v$ for each vehicle $v$ in the fleet. AMoDSim is able to simulate on-line optimization algorithms, in which the schedules are continuously modified. To avoid violating some user constraints, the feasibility should be checked at any modification. For example, suppose we modify $S_v$ by inserting a new stop-point $sp=(q,t,\Delta t)$ at position $k$, obtaining a new schedule $S_v^{(k)}=[sp_0,\dots,sp_{k-1},sp,sp_{k},\dots,sp_n]$. The detour the vehicle does to serve $sp$ determines an additional delay on all the stop-points after the $k$-th. If we denote with $\hat t_i^{(k)}$ the estimated stop-point time of $sp_i$ after the insertion, the additional delay is $\Delta d_i^{(k)} \equiv \hat t_i^{(k)}- \hat t_i$ and it is easy to show that:
\[
	\Delta d_i^{(k)} = 
    \begin{cases}
    0, & \textnormal{if }i<k \\
    \tau(q_{k-1},q) + b + \tau(q,q_k) - \tau(q_{k-1},q_k), & \textnormal{if }i\ge k
    \end{cases}
\]
where $b$ is the time for alighting or boarding related to $sp$.
To check whether the modified schedule is feasible, not only must we check that the time constraints of the new $sp$ are satisfied, but also that the time constrains are satisfied for all the stop-points already present in the schedules, i.e., $d_i+\Delta d_i^{(k)}\le \Delta t_i$ for $i=0,\dots,n$.

\subsection{Examples of optimization strategies}
\secL{strategies}
To give a more concrete idea of the model we discussed in the previous section, we now describe two heuristics we implemented in AMoDSim and some possible assumptions about the request constraints expressed by users. We adopt such heuristics and assumptions in the case study of Sec. \ref{sec:case-study}. However, we emphasize that the simulator is more general and can be used in different ways.

Recall a request sent by a user is composed by a stop-point $sp=(q,t,\Delta t)$ for the pick-up and another $sp'=(q',t',\Delta t')$ for the drop-off. We assume that the user would like to be picked-up immediately, i.e., $t=t_\textnormal{now}$ and to be dropped-off as in the ideal case in which a vehicle is immediately at her disposal and can bring her to the destination in the shortest path, without detours, i.e., $t'=t_\textnormal{now}+\tau(q,q')$.

We implement two optimization strategies, namely Radio-Taxi and Insertion Heuristic. With the former each vehicle can serve one passenger at a time, while the latter allows ride sharing, i.e., the same vehicle can serve multiple passengers at a time.

%\rev{AD}{}{Anche nel caso del RadioTaxi costruiamo lo schedule per ogni veicolo: ad ogni nuova richiesta, costruiamo lo schedule mettendo questa in coda alla lista di stop point di ogni veicolo (o in testa se un veicolo non ha sp). Ogni schedule avrà quindi anche in questo caso un costo: -se il veicolo è vuoto, il costo è dato dal tempo necessario a raggiungere il pickup dalla posizione attuale del veicolo + il tempo per andare dal pickup al dropoff; - se il veicolo ha altri stop point il costo è dato dal tempo necessario per andare dall'ultimo stop point al nuovo pickup + il tempo per andare adl pickup al dropoff.  Il coord anche in questo caso assegna la richiesta al veicolo che ha costo minore.} \\
We first describe the \emph{Insertion Heuristic}, loosely inspired by ~\cite{insertionheuristic}. 
%\rev{AD}{}{Nel paper, l'algoritmo di insertion heuristic è descritto in modo generalizzato e questa generalizzazione include anche la nostra descrizione. Ad esempio il costo di uno schedule lo considera come somma tra i waiting time e i travel time, mentre noi lo consideriamo come somma dei tempi necessari a raggiungere tutti gli stop point. Questo porta cmq ad avere lo stesso risultato. Ho però notato una differenza che secondo me è rilevante: loro parlano di max waiting time e di un fattore che chiamano detour, che se ho capito bene indica il numero massimo di deviazioni ammesse da un passeggero dopo essere salito a bordo. Nelle loro sim lo pongono pari a 2 e dicono: "the demand impacts need to be investigated within the given detour constraint." Noi invece non poniamo alcun vincolo sulle deviazioni che possono essere fatte perche ci riferiamo al maxDelay.} 
%
The cost function $c(S_v)$ is chosen in order to represent the user experience. More precisely, the cost of a schedule $S_v=[sp_0,\dots,sp_n]$ is the sum of the estimated delays $d_i$, as defined in Sec. \ref{sec:time}, of all its stop-points, i.e., $c(S_v)=\sum_{i=0}^n d_i$. The Insertion Heuristic attempts to minimize the marginal cost when serving an additional request. Suppose a new request is sent, consisting of the stop-points $sp,sp'$ for the pick-up and drop-off, respectively. Assigning the new request to any vehicle, will increase the cost of its schedule, i.e., the sum of the delays suffered by its stop-points. Let us take any vehicle $v$ and denote with $S_v^{(k,k')}$ the schedule obtained from $S_v$ by inserting the pick-up $sp$ in the $k$-th position and the drop-off $sp'$ in the $k'$-th position, with $k'>k$. If the modified schedule is infeasible, we set $c(S_v^{(k,k')}=\infty)$. We compute the best placement of drop-off and pick-up, which minimizes this increase in cost, i.e., 
\[
(k^v, k'^v) = \arg\min_{(k,k'), k'>k} \left( c(S_v^{(k,k')}) - c(S_v) \right)
\]
We repeat the same computation for all the vehicles and we choose the one whose marginal cost is minimum, i.e.:
\[
v^* = \arg\min_v \left( c(S_v^{(k^v,k'^v)}) - c(S_v) \right)
\]
Finally, we assign the request to vehicle $v^*$ and place the pick-up and drop-off in the $k^{v*}$-th and $k'^{v*}$-th positions, respectively. 

The \emph{Radio-Taxi} strategy is a constrained version of Insertion Heuristic, in that we impose that each pick-up be followed in any schedule by the correspondent drop-off, which ensures that at most one passenger is in the vehicle at any moment.

\subsection{Vehicle Movement}
\label{sec:movement}
All vehicles travel through the links of the network, i.e., roads, at a predefined \emph{cruising speed}. Each link has a length, which determines the time needed to traverse it.
Obviously, when a vehicle alternates between a stop-point and another, its speed does not go from 0 to the cruising speed and back to 0 instantaneously.
Therefore, we introduce a parameter $t_a$ ($t_d$), which represents the time lost for accelerating (decelerating). When a vehicle reaches a stop-point $sp_i$, we keep it in that node for an additional time $b_i+t_a+t_d$, before sending it again to the link toward the next stop-point.

\section{Software Architecture}
\label{sec:simulator}
AMoDSim is a simulation platform developed on top of Omnetpp\cite{omnet}. It is designed to be configurable, modular, event-based, algorithm-oriented and extensible with custom optimization strategies and network topologies.\\
The simulator models the road network as a set of nodes, i.e., geographical locations that could be origins and destinations of the service requests, connected through links, i.e., road connections between different locations. A \emph{vehicle} is represented as a \emph{packet} traveling through the links. A node is a compound-module composed of three sub-modules: queue, routing and application. A node has one queue module per each outgoing or incoming link. Each \emph{Queue} module forwards (receives) packets to one of the outgoing links (from one of the incoming links). The \emph{Routing} module (i) decides to which of the outgoing links a packet should be forwarded and (ii) checks, every time a vehicle passes, whether the node is one of its stop-points, in which case the vehicle is passed to the Application module. The \emph{Application} module implements multiple functions:
\begin{itemize}
\item It generates user requests, as pairs of stop-points (one for the pick-up and one for the drop-offs). The generation obeys to a pre-determined stochastic process. So far, Poisson arrivals are implemented.
\item It receives all the vehicles for which the node in question is a stop-point, checks the next stop-point, accessing a data-structure storing all the schedules and sends the vehicle to it. At the same time, it also notifies the coordinator, so that it can update the schedule in question.
\item It keeps the vehicles that are idling at the node with an empty schedule. In this case, it also receives a signal from the coordinator if a new schedule is assigned to the idling vehicles and sends them to their new stop-point.
\end{itemize}
\begin{figure}[H]
  \centering
	\includegraphics[scale=0.45]{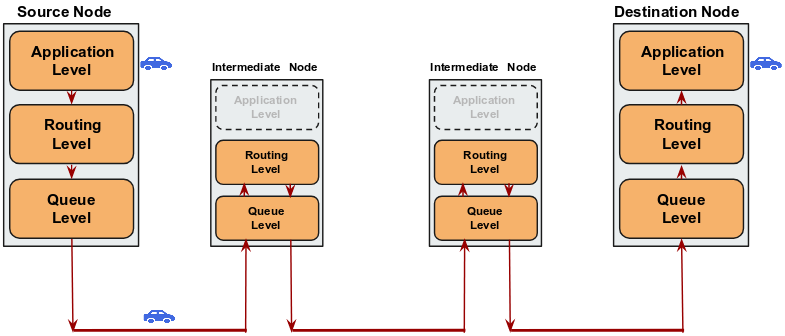}
  \setlength{\belowcaptionskip}{-20pt}
	\caption{A \emph{trip} example}
\end{figure}

The \emph{Coordinator} manages the incoming trip requests, implements the trip allocation strategies and assigns each request to a vehicle, according to the implemented optimization strategy. It has been designed to be easily extensible with custom allocation strategies. 
We implemented a modular Coordinator within a hierarchical structure where the superclass implements the standard functions. One can extend such superclass and implement the logic of her matching algorithm.

%\begin{figure}[H]
%  \centering
%	\includegraphics[scale=0.2]{network}
%	\caption{AMoD network}
%\end{figure}

\subsection{AMoD Performance Metrics}
AMoDSim collects data during its execution and produces a set of results that enable statistical analysis related to both the point of view of the provider and of users.

Regarding the provider viewpoint, AMoDSim provides the following information per-vehicle: (i) distance traveled, (ii) number of passengers on board, (iii) requests picked-up but not yet dropped-off, (iv) number of pick-ups already in the schedule but not yet completed, (v) total requests assigned, (vi) the time the vehicle has spent idle or with $p$ passengers, where $p$ ranges from 1 to the number of per-vehicle seats.

Moreover, for each of the collected metric, AMoDSim computes aggregated fleet statistics, as sum, minimum, maximum, mean, standard deviation, median and 95th percentile.

At each time frame, the following information about the users' requests received up to that time are collected: (i) length of the submitted requests, (ii) number of requests that the system has received and assigned to the vehicles, (iii) number of requests that the system has rejected because it could not serve them within the time constraints, (iv) number of requests that the system is processing at the snapshot time.

The level of quality for the users is described by the following per-user quantities: (i) time that users spent in the pick-up location waiting for the vehicle, (ii) actual time that the user spent in the vehicle, (iii) \emph{Stretch}, i.e., the ratio between the actual trip time and the preferred one, which is the time between the preferred pick-up and drop-off times.

\section{Case Study}
\label{sec:case-study}
We showcase the capabilities of AMoD in a simple case study, in which we launched a campaign of 1800 simulations.
We compare the performance of the Insertion Heuristic to the Radio-Taxi. We show how AMoDSim allows to find interesting insights on the AMoD systems and answer questions like: what is the fleet size needed to sustain a certain request rate? Which kind of vehicles should be employed (of how many seats)? What is the sharing level, i.e., how effectively are we able to condensate different user rides in few vehicle schedules? By how much sharing rides allows to reduce the fleet size needed? How efficient is vehicle usage, e.g., how much time vehicles are idle? We underline that the findings we get are not necessarily general properties of every AMoD systems, but depend on the particular optimization strategy we adopt and the particular scenario. Therefore, our goal is to show how other researchers can obtain similar findings with AMoDSim about their strategies and their scenarios. Finally, we show the computational performance of AMoDSim. We are aware that the quality of AMoDSim cannot be validated only by the  case study we present here. Part of our future work is to apply AMoDSim to different scenarios and to validate by comparing it with other simulators. This latter point requires careful thinking, since other simulators are not directly comparable, for the reasons discussed in \secR{introduction}. We also believe that the best way to make AMoDSim reach full maturity is its adoption by other researchers for their studies, which would help in understanding and improving its limits.

\subsection{Scenario}
We use Manhattan Grid that covers an area of $60\textit{km}^2$, equivalent to Manhattan, with static link travel times as in \cite{Mahmassani2018}. We consider different configurations of the fleet of vehicles to study the performance of multiple ride-sharing degrees and fleet size. We perform simulations starting from single-seater up to 10-seater minibus and a fleet of 500 up to 9000 vehicles. We assume a cruising speed of $35\textit{kmph}$ and a constant acceleration and deceleration of $1.676$ mpss, resulting in a $t_a+t_d=11.5$ (see Sec.\ref{sec:movement}) as in \cite{timeloststop}.    
Thus, the vehicles have a constant acceleration (deceleration) of $1.676$mpss ($-1.676$mpss). Users submit requests with Poissonian arrivals as in \cite{Jung2013} with rate ranging from 20 up to 640 requests per hour per $\textit{km}^2$ compatible with the scenarios employed in the literature \cite{Alonso-Mora2017,insertionheuristic}. As for the $b_i$ of a pick-up (drop-off) stop point $sp_i$, i.e. the time need for boarding (alighting), we assume 5 seconds (10 seconds) as in \cite{board_alig_time}. All results are collected running 4h simulations.

\subsection{Results}
In this section, we first give an example of analysis possible in AMoDSim and then discuss its computational performance.
\subsubsection{Sharing opportunities for an AMoD provider. }
\begin{figure}[h]
\begin{subfigure}[b]{.32\linewidth}
\includegraphics[width=\linewidth]{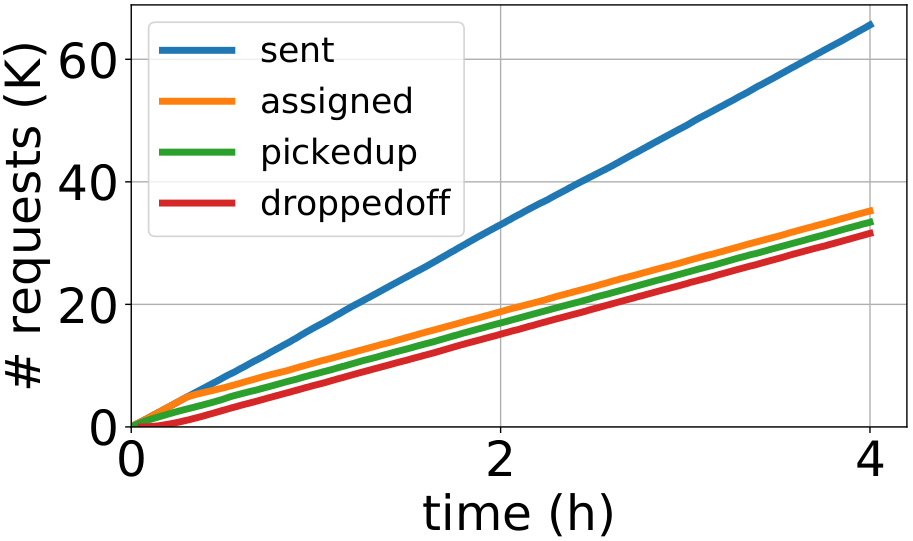}
\end{subfigure}
\begin{subfigure}[b]{.32\linewidth}
\includegraphics[width=\linewidth]{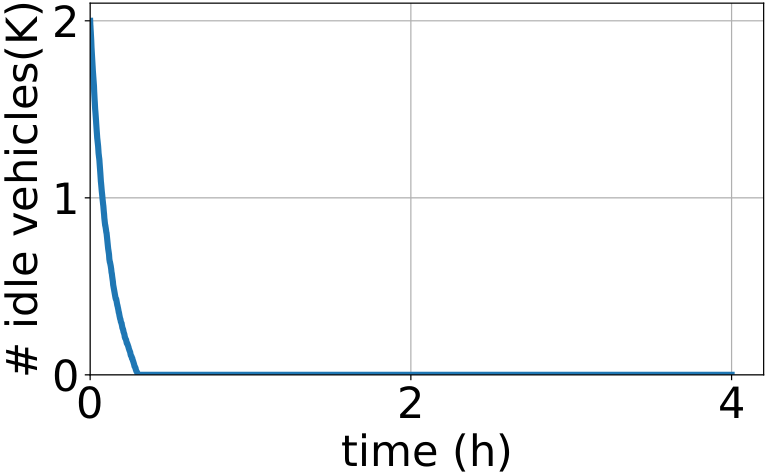}
\end{subfigure}
\begin{subfigure}[b]{.32\linewidth}
\includegraphics[width=\linewidth]{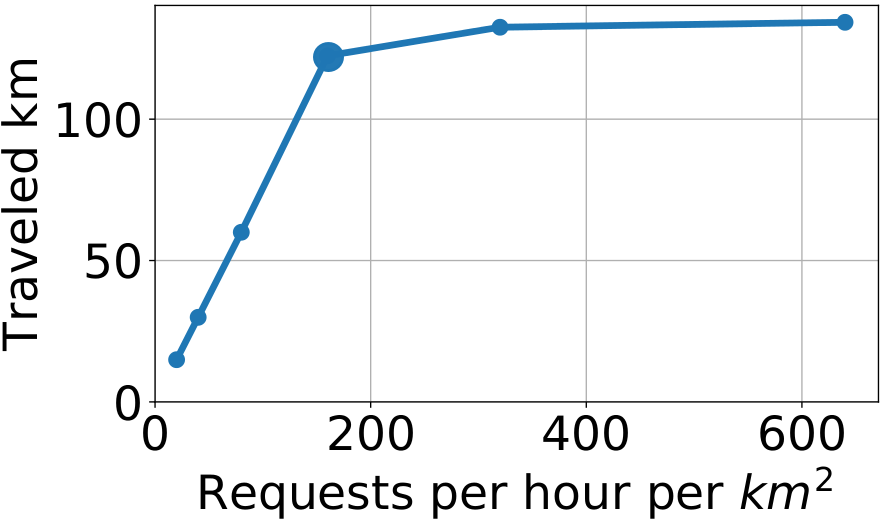}
\end{subfigure}
\caption{RadioTaxi: maximum extra-time $\Delta t=15$min, 2K vehicles. In the left and middle figure, the rate is 320req/h/$\textit{Km}^2$}
\label{fig:RequestsPerTimeRadio}
\end{figure}

\begin{figure}[h]
\begin{subfigure}[b]{.32\linewidth}
\includegraphics[width=\linewidth]{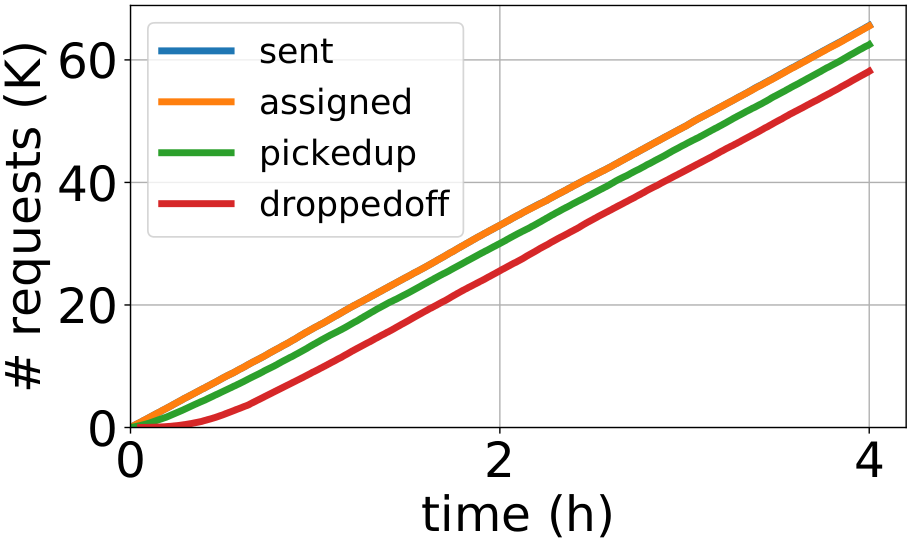}
\end{subfigure}
\begin{subfigure}[b]{.32\linewidth}
\includegraphics[width=\linewidth]{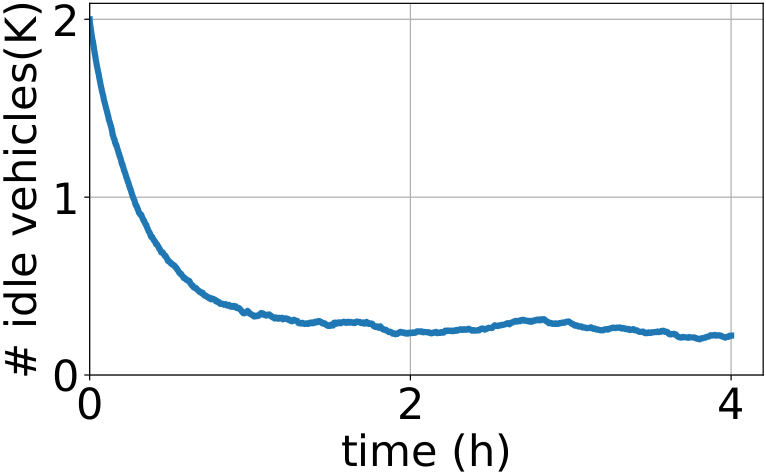}
\end{subfigure}
\begin{subfigure}[b]{.32\linewidth}
\includegraphics[width=\linewidth]{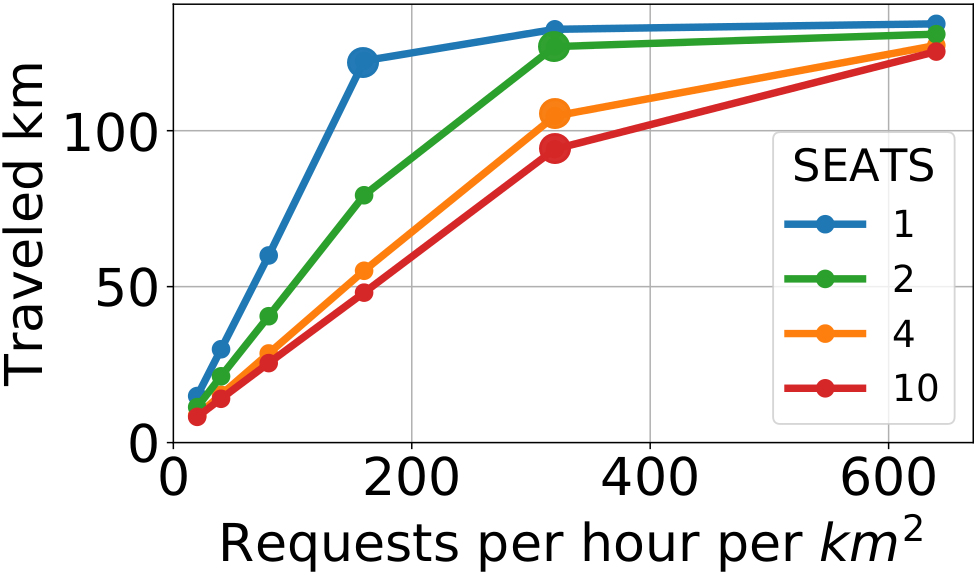}
\end{subfigure}
\caption{Insertion Heuristic: $\Delta t=15$min. In the left and middle plots, 320req/h/$\textit{Km}^2$ and 2K 4-seater vehicles are used.}
\label{fig:RequestsPerTimeHeur}
\end{figure}

We investigate the factors determining the \emph{sharing degree} and its impact on the provider and the users. The sharing degree is the capacity of an AMoD provider to exploit the fact that a single resource (vehicle) can be used to serve multiple requests. This concept, at the core of the sharing economy, cannot be quantified in a single value, but emerges from a set of different indicators that we discuss here.
Fig. \ref{fig:RequestsPerTimeRadio} shows the performance of Radio-Taxi. It is clear that the system is saturated: only 35K requests are served over 65K and the number of idle vehicles goes down to zero in few minutes. Fig. \ref{fig:RequestsPerTimeHeur} shows that under the same conditions, Insertion Heurisitc with a fleet of 4-seater 2K vehicles allows to meet all the requests. Observe also that the total number of kilometers traveled, a proxy for the provider cost, decreases considerably by increasing the number of seats, since the sharing opportunities increase.


\begin{figure}[h]
\begin{center}
\begin{subfigure}[b]{.25\linewidth}
\includegraphics[width=\linewidth]{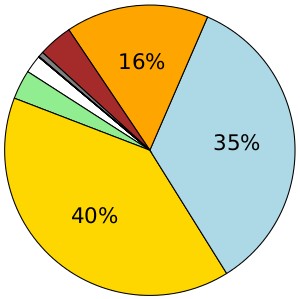}
\caption{$\Delta t=5$min} \label{fig:ttime_delay5}
\end{subfigure}
\begin{subfigure}[b]{.25\linewidth}
\includegraphics[width=\linewidth]{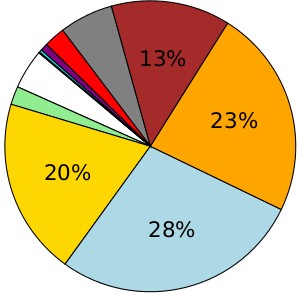}
\caption{$\Delta t=15$min} \label{fig:ttime_delay15}
\end{subfigure}
\begin{subfigure}[b]{.25\linewidth}
\includegraphics[width=\linewidth]{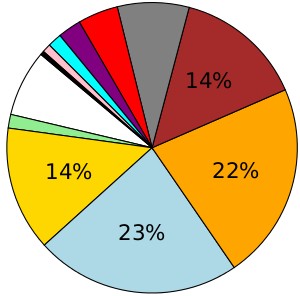}
\caption{$\Delta t=30$min} \label{fig:ttime_delay30}
\end{subfigure}
\end{center}
\medskip

\begin{subfigure}[b]{\linewidth}
\includegraphics[width=\linewidth]{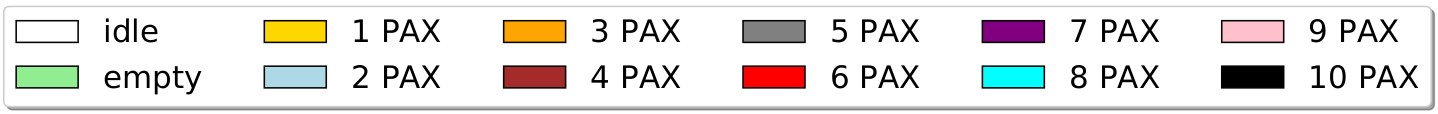}
\end{subfigure}
\setlength{\belowcaptionskip}{-20pt}
\caption{Vehicle occupancy with 1K 10-seater vehicles and a rate of requests 320 per hour per $\textit{km}^2$} 
\label{fig:occupancy}
\end{figure}

The sharing degree is well summarized by Fig.\ref{fig:occupancy}, which shows the fraction of time vehicle spend, on average, with 0 (idle), 1, 2, ... passengers. Intuitively, if we allow users to express a tight extra-time constraint $\Delta t$, the sharing opportunities shrink and we can just afford few passengers at a time, in order to meet the constraints of all of them.

Note that, even with a long $\Delta t$, more than 6 seats are rarely utilized. This suggests that, if we want to implement a minibus-like service, strategies different from Insertion Heuristic must be used (which is an interesting subject to investigate). Observe also that high capacity vehicles would be fully utilized only if $\Delta t$ is too tight.
In other words the type of vehicles to be used depends on the type of service that the provider wishes to offer and the level of service users expect.

\subsubsection{Mean waiting time. }
\begin{figure}[h]
\begin{center}
\begin{subfigure}[b]{.32\linewidth}
\includegraphics[width=\linewidth]{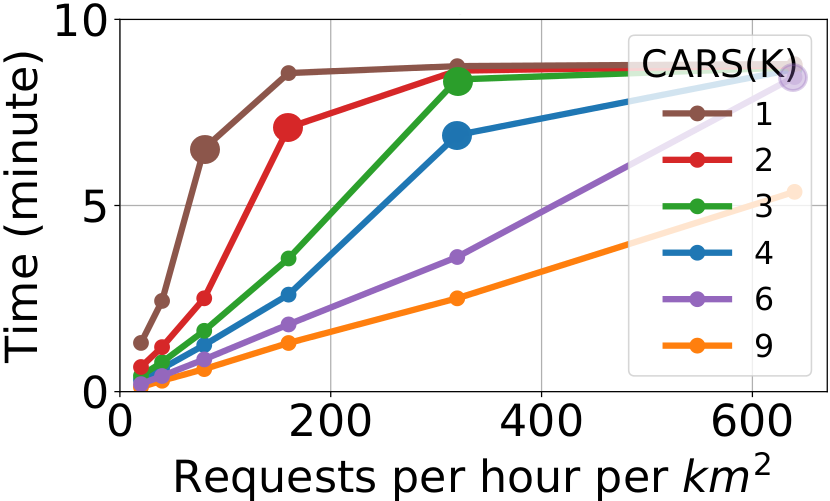}
\caption{Radio-Taxi} \label{fig:wtime_radiot}
\end{subfigure}
\begin{subfigure}[b]{.32\linewidth}
\includegraphics[width=\linewidth]{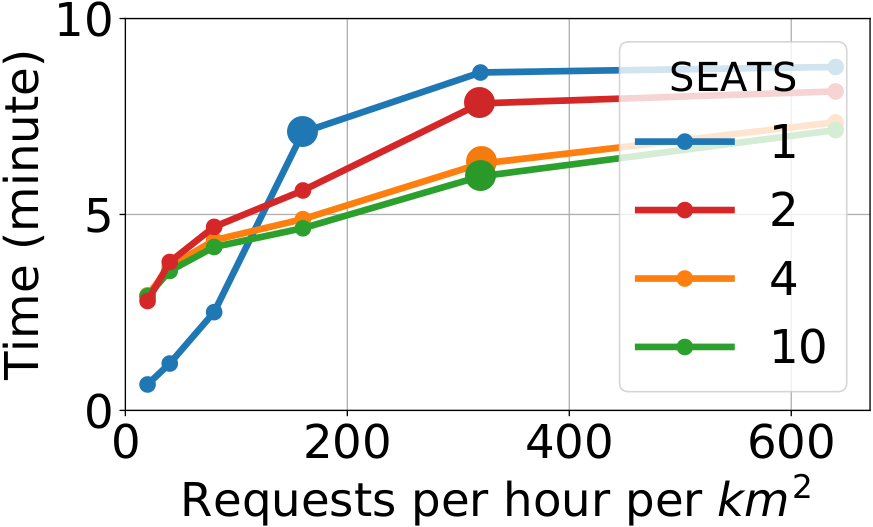}
\caption{Heuristic: vehicles=2K} \label{fig:wtime_heuristic}
\end{subfigure}
\end{center}
\setlength{\belowcaptionskip}{-20pt}
\caption{Mean waiting time with a maximum delay $\Delta t=10$ minutes. }
\label{fig:waitingtime}
\end{figure}

In a RadioTaxi-based AMoD system, the only way to serve a higher service demand is to increase the fleet size. Moreover, a large fleet reduces the Waiting Time (WT), which is shown in Fig. \ref{fig:wtime_radiot}. With the Insertion Heuristic another parameter impacts the user experience, namely the vehicle seats. In Fig. \ref{fig:wtime_heuristic} we use large points to indicate the first value of request rate in which we observed the system is in saturation, i.e., it is not able to serve all the requests, e.g., Fig.\ref{fig:RequestsPerTimeRadio}. 
Observe that when the system is not saturated, the best WT are measured with 1 seater vehicles, since each is dedicated entirely to a single user each time and the user does not make detours due to sharing with others. However, the system saturates at only 160req/h/$\textit{km}^2$. On the contrary, larger vehicles allow to serve a more intense demand without saturation, which translates in a better WT for the users.

\subsubsection{Computation time and memory consumption. }
\begin{figure}[h]
\begin{subfigure}[b]{.32\linewidth}
\includegraphics[width=\linewidth]{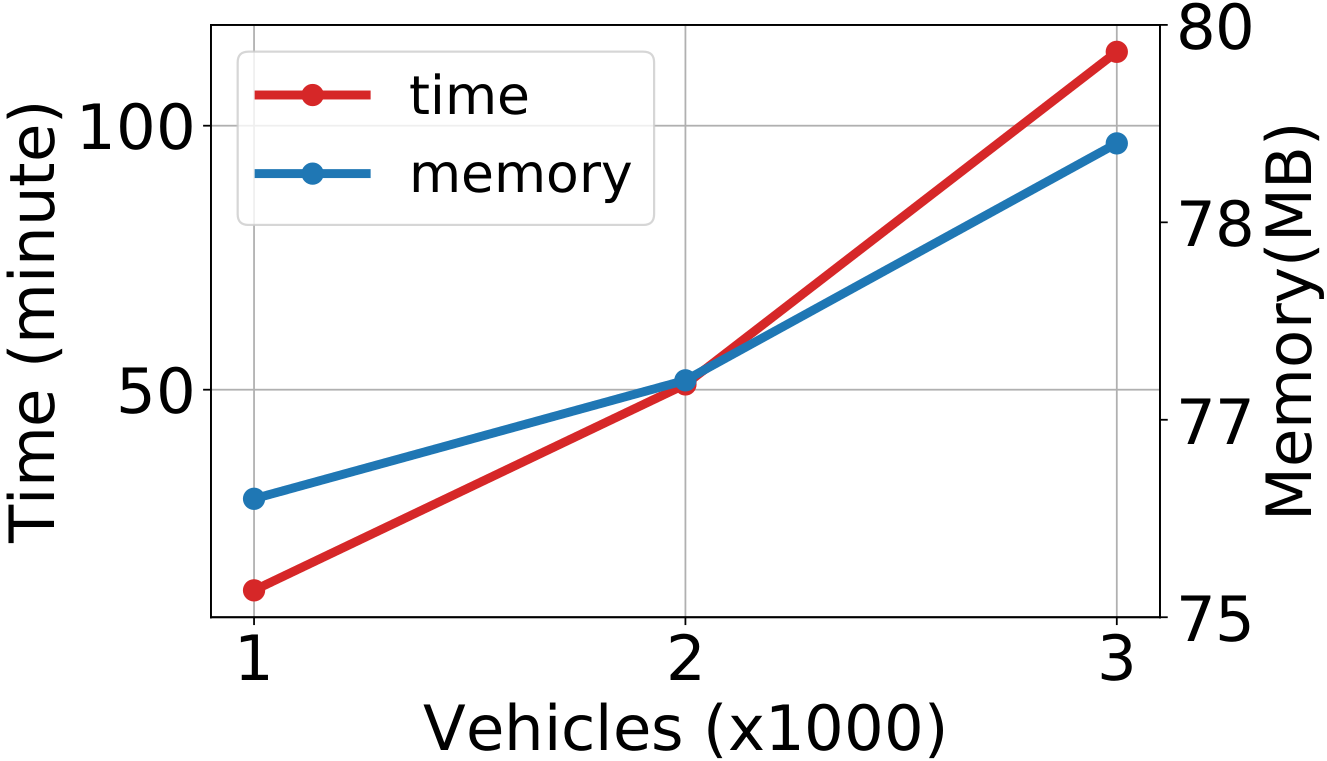}
\caption{rate=160, seater=4} \label{fig:perf_vehic}
\end{subfigure}
\begin{subfigure}[b]{.32\linewidth}
\includegraphics[width=\linewidth]{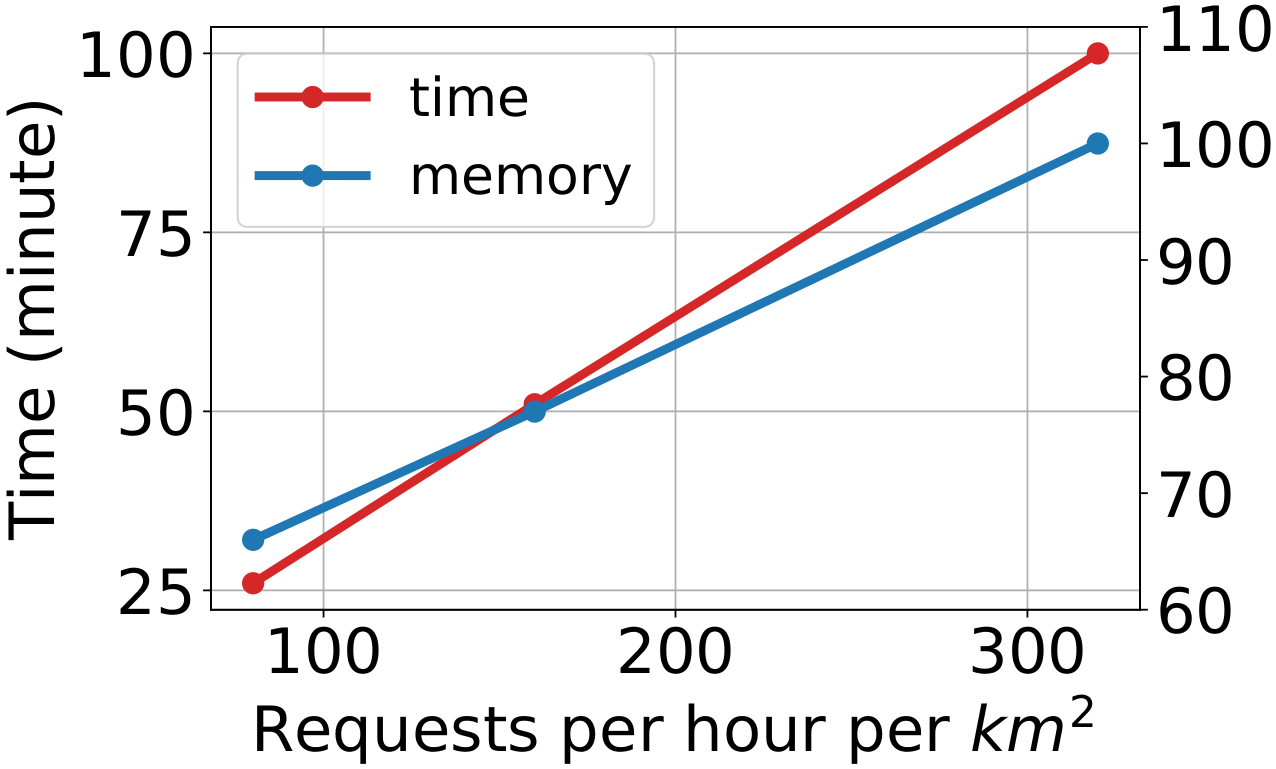}
\caption{vehicles=2K, seater=4} \label{fig:perf_rate}
\end{subfigure}
\begin{subfigure}[b]{.32\linewidth}
\includegraphics[width=\linewidth]{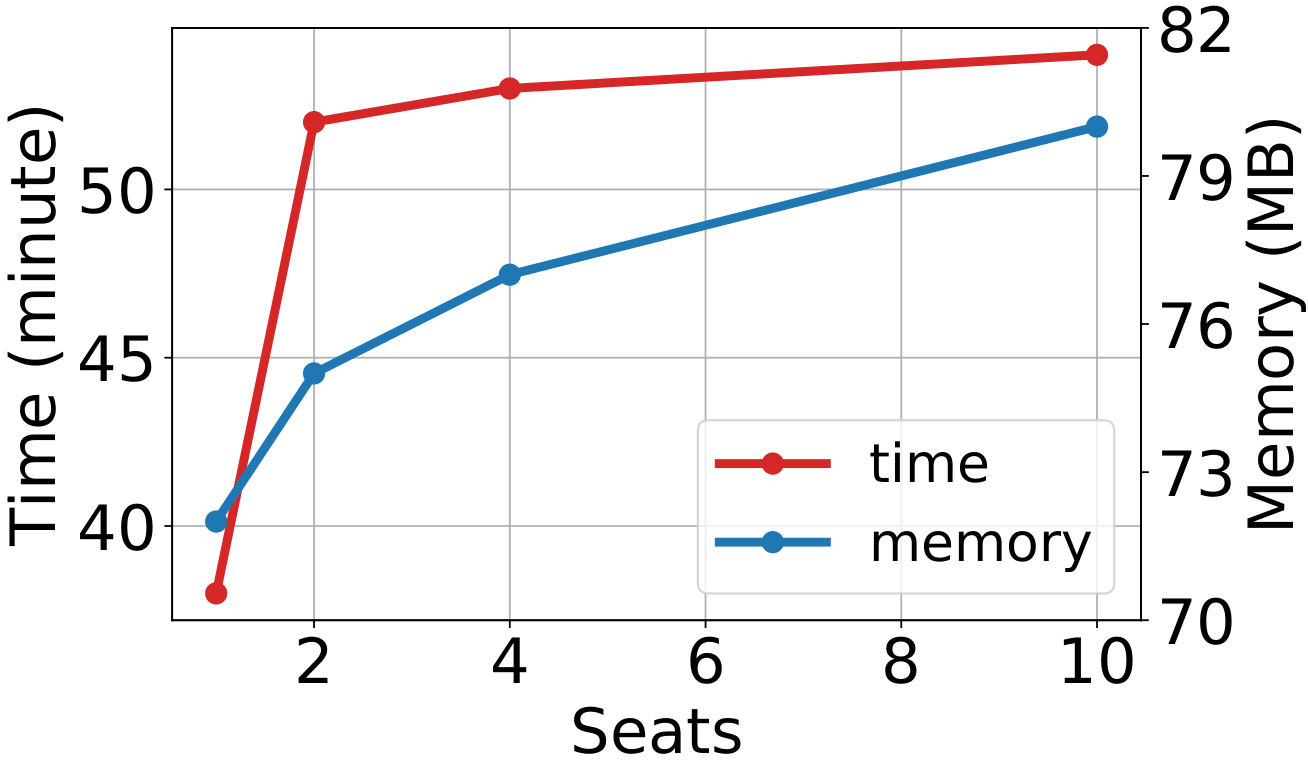}
\caption{rate=160, vehicles=2K} \label{fig:perf_seats}
\end{subfigure}
\setlength{\belowcaptionskip}{-20pt}
\caption{Computation time and Memory consumption: $\Delta t=15$min}
\label{fig:performance}
\end{figure}

In this section we discuss the single-run computation time and the peak memory consumption of AMoDSim, which we observed in our case study. Note that comparison with other simulators is not possible here for the reasons discussed in Sec.\ref{sec:related}: the case-specific simulators are not available and the transportation simulators are out of scope and would have required input data that do not exist for the scenarios considered. Fig. \ref{fig:perf_vehic} and \ref{fig:perf_rate} show how both the computation time and the memory consumption grow with the number of vehicles and the rate of requests, as expected. Fig. \ref{fig:perf_seats} shows how the increase in computation time is significant moving from single-seater to 2-seater vehicles and is low moving from 4-seater to 10-seater. This may be due to the fact that vehicles spend most of the time with no more than 4 passengers anyway (Fig.\ref{fig:occupancy}).

\section{Conclusion}
NOT IN THIS DRAFT

\section{Acknowledgement}
NOT IN THIS DRAFT.

\bibliographystyle{plain}
\bibliography{sample2}
\end{document}